\def\keV{{\rm keV}}

\def\rmp{Rev. Mod. Phys.}

\def\Omat{\Omega_{\rm M}}
\def\Olam{\Omega_{\Lambda}}
\def\Obar{\Omega_{\rm b}}
\def\Lx{L_{\rm X}}
\def\Tx{T_{\rm X}}

\def\gtrsim{\stackrel{>}{\sim}}
\def\lesssim{\stackrel{<}{\sim}}

\def\mnras{MNRAS}
\def\apj{ApJ}

\def\aap{A\&A}
\def\nat{Nature}
\def\apjl{ApJ}

\documentclass[useAMS,usenatbib]{mn2e}
\usepackage{graphicx}
\usepackage{epsfig}
\title[The baseline intracluster entropy profile from gravitational structure formation]{The baseline intracluster entropy profile from gravitational structure formation}
\author[G. Mark Voit. Scott T. Kay,  and Greg L. Bryan]{G. Mark Voit$^{1}$\thanks{E-mail:
voit@pa.msu.edu (GMV); skay@astro.ox.ac.uk (STK); gbryan@astro.columbia.edu (GLB)},  Scott T. Kay$^{2}$, and Greg L. Bryan$^{3}$ \\
$^{1}$Department of Physics and Astronomy, 
	  Michigan State University, 
	  East Lansing, MI 48824, USA \\
$^{2}$Physics Department,
            University of Oxford, 
            Keble Road, 
            Oxford OX1 3RH, UK \\
$^{3}$Department of Astronomy,
            Columbia University,
            New York, NY  10027, USA}

\begin{document}

\date{Received 2005 May 25, revised 2005 September 16, accepted 2005 September 21}


\maketitle

\begin{abstract}
The radial entropy profile of the hot gas in clusters of galaxies tends to follow a power law in radius outside of the cluster core.  Here we present a simple formula giving both the normalization and slope for the power-law entropy profiles of clusters that form in the absence of non-gravitational processes such as radiative cooling and subsequent feedback.  It is based on seventy-one clusters drawn from four separate cosmological simulations, two using smoothed-particle hydrodynamics (SPH) and two using adaptive-mesh refinement (AMR),  and can be used as a baseline for assessing the impact of non-gravitational processes on the intracluster medium outside of cluster cores.  All the simulations produce clusters with self-similar structure in which the normalization of the entropy profile scales linearly with cluster temperature, and these profiles are in excellent agreement outside of $0.2 r_{200}$. Because the observed entropy profiles of clusters do not scale linearly with temperature, our models confirm that non-gravitational processes are necessary to break the self-similarity seen in the simulations.  However, the core entropy levels found by the two codes used here significantly differ, with the AMR code producing nearly twice as much entropy at the centre of a cluster.
\end{abstract}


\setcounter{footnote}{0}

\section{Introduction}

Purely gravitational structure formation ought to produce clusters of galaxies
with nearly self-similar structure, whose X-ray luminosity $\Lx$ scales with 
temperature $\Tx$  as $\Lx \propto \Tx^2$ 
\citep{kaiser86}.  Clusters created in hydrodynamical simulations 
with cosmological initial conditions indeed follow this scaling relation
\citep[e.g.,][]{nfw95, enf98}, but observed clusters do not.  Instead they follow a relation 
closer to $\Lx \propto \Tx^{2.8}$ \citep{es91, djf95, mark98, af98, ae99}.  
Somehow the non-gravitational cooling and heating processes associated 
with galaxy formation intervene to break the expected self-similarity, with 
consequences that are more severe in low-temperature clusters than in 
high-temperature clusters \citep[][see Voit 2005 for a recent review]{kaiser91, eh91}. 

The scaling behavior of the $\Lx$-$\Tx$ relation can largely be understood
in terms of radiative cooling and the feedback it triggers \citep{vb01}.  
Gas that can cool within a Hubble time must either condense, forming stars or 
cold baryonic clouds, or it must be reheated somehow, probably by 
supernovae or AGN activity triggered by the condensing gas.  Because $\sim 85$\% 
of the baryons associated with a massive cluster appear to be in the hot intracluster 
medium (ICM), a certain amount of supernova or AGN feedback seems necessary 
to prevent too many of the baryons from condensing during the formation
of a cluster's galaxies \citep{wf91, bpbk01}.   However, the $\Lx$-$\Tx$ relation itself is
relatively insensitive to the total amount of feedback energy, as long as the 
energy input into the reheated gas is sufficient to keep it from cooling again 
\citep{borg02, ktt03, tbsmmm03, Vald03}. Thus, one needs to look beyond
the $\Lx$-$\Tx$ relation in order to assess the thermodynamic impact
of supernovae and AGNs on the state of the ICM.

One good place to look for more information is in the spatially-resolved entropy
profiles of clusters and groups, which preserve a record of the cooling and
heating processes responsible for similarity breaking in clusters \citep{vbbb02, vbblb03,
Kay04, ktjp04}.  Specific entropy, represented in this paper by
the quantity $K = Tn_e^{-2/3}$, where $n_e$ is the electron density,
is more closely tied to the thermodynamic history of a cluster than is 
temperature, because the thermal energy of heated gas can be converted 
into gravitational potential energy as the heated gas expands in the confining 
potential well.\footnote{The classical thermodynamic entropy per particle is $s = \ln K^{3/2} +
{\rm const.}$ in an ideal monoatomic gas.}  Under certain circumstances, a large
amount of energy input produces only a small rise in the 
luminosity-weighted temperature $\Tx$ \citep[e.g.,][]{vbbb02}.
The specific entropy of the ICM, on the other hand, always rises when heat energy
is introduced and always falls when radiative cooling carries heat energy away.  

The entropy profiles of clusters and groups can now be measured out to a significant
fraction of the scale radius $r_{200}$, within which the mean mass density is 200
times the critical density.
Those measurements show that entropy levels in the cores of clusters, where
the $\Lx$-$\Tx$ relation is determined, scale as $K_{0.1} \equiv K(0.1 r_{200}) 
\propto \Tx^{2/3}$ \citep{psf03}, as expected if radiative cooling and
associated feedback govern the core entropy level \citep{vp03}.
More surprisingly, entropy measurements at larger cluster radii are hinting
that this scaling relation applies to the entire entropy profile.  Deep {\em XMM-Newton}
observations of five clusters whose temperatures span a range of $\sim$3.5
show that the scaled profile $\Tx^{-2/3} K(r/r_{200})$ is independent of cluster 
temperature \citep{pa03, pa04}.  Likewise, an analysis of lower-quality data 
on a larger number of clusters also suggests that $K(r/r_{200}) \propto \Tx^{2/3}$
at the scale radius $r_{500}$, within which the mean matter density is 500 times the
critical density \citep{psf03}.

Rather than totally breaking the self-similarity of clusters, galaxy formation appears to 
alter the power-law scaling of $K(r/r_{200})$ with $\Tx$
without appreciably changing the overall shape of the entropy profile.
Exactly how heating and cooling would conspire to produce such a shape-preserving
shift in the normalization of an intracluster entropy profile is unknown.  One possibility involves
smoothing of the intergalactic medium by supernovae or AGN energy input prior to accretion, 
which lowers the mass-weighted mean density $\bar{\rho}_{\rm acc}$ of the infalling gas.  
Because the amount of entropy generated in that gas when it passes through accretion 
shocks of velocity $v_{\rm acc}$ is $K_{\rm acc} \sim v_{\rm acc}^2 / \bar{\rho}_{\rm  acc}^{2/3}$, 
smoothing of gas that would otherwise be bound to accreting subhalos boosts the 
post-accretion entropy level of the ICM \citep{vbblb03, psf03, Borgani05}.  Another 
possibility, illustrated in simulations by \citet{Kay04}, involves uncompensated cooling, 
which allows high-entropy gas to sink  to smaller radii as the core gas condenses,
but pure cooling does not appear to reproduce the $K \propto \Tx^{2/3}$
relation at $r_{500}$. 

A proper analysis of the observations to determine the true source of the entropy boost
requires knowing what the baseline entropy profile of a cluster would be like in the 
absence of galaxy formation.  To that end, this paper compares the results of four 
different hydrodynamical simulations of purely gravitational cluster formation 
with the aim of deriving a simple analytical formula for that baseline self-similar 
entropy profile.  The hydrodynamical computations in the simulations employ entirely 
different numerical algorithms: some use smoothed-particle hydrodynamics (SPH) 
while the others use adaptive-mesh refinement (AMR).  Thus, to the extent
that these two techniques produce convergent results, our comparison provides 
a reliable baseline profile with which to interpret the observations.
Our comparison shows that the codes agree and are presumably reliable outside the cores of clusters, where entropy levels are relatively high, but disagree inside the cores, where entropy levels are lower.  This discrepancy is not yet understood and may result from differences in how the codes treat small-scale shocks and mixing processes.

While the cosmological parameters used to specify the initial conditions in the
simulations are similar---$\Lambda$CDM, with matter density $\Omat \approx 0.3$, 
dark-energy density $\Olam \approx 0.7$, baryon density $\Obar \approx 0.04$,
Hubble constant\footnote{We define $h \equiv H_0/(100 \, {\rm km \, s^{-1} \, Mpc^{-1}})$.}
$h \approx 0.7$ and power-spectrum normalization $\sigma_8 \approx 0.9$---the 
initial conditions themselves are not identical.  The  
simulations do not model the evolution of the same field of density perturbations.  
Instead, we compare results for many different clusters within
representative but not identical volumes of the universe, as modeled by each
code.

The paper proceeds as follows.  Section 2 establishes a non-dimensional framework 
for comparing the entropy profiles of clusters having different masses then describes 
the two codes we use and presents results from each code.  In each case we determine 
the median dimensionless entropy profile and the scatter about the median.
Knowing the variance in entropy owing to the non-steady nature of merging should help 
observers establish whether additional variance generated by the stochasticity
of non-gravitational processes is needed to explain the observations.  
Section 3 compares the results from the two codes and provides 
a simple analytical form for the baseline entropy profile outside the cluster core
that adequately represents the results of both simulations.  
Because the entropy profiles determined with the two
codes disagree somewhat in the cluster core, the formula is valid only for
$0.2 \lesssim r/r_{200} \lesssim 1.0$.   Section 4 summarizes our results.

\section{Simulations of Non-Radiative Clusters}

Clusters that form without radiating away any of their thermal energy are unphysical,
but they constitute a useful baseline against which to measure the effects of radiative
cooling and non-gravitational heating \citep[e.g.,][]{vbbb02, f99_sb}.  In spite of the obvious fact 
that accretion shocks generate an enormous amount of entropy during cluster formation.
Such clusters are sometimes called ``adiabatic" clusters in the literature on this 
subject.  Because we prefer to reserve the term ``adiabatic'' for isentropic processes 
that do not involve shock heating, we will refer to these clusters as ``non-radiative" clusters 
\citep[as in][]{mtkpc01}.

This section examines two populations of non-radiative clusters generated using both 
SPH and AMR techniques.  The properties of these simulated 
clusters are most easily compared if we scale away the dependence of dimensional 
quantities on halo mass, a procedure that would lead to identical temperature, 
density, and entropy profiles for each cluster if clusters were precisely self-similar.  
Thus, we begin by outlining the scaling behavior expected in the self-similar case and then
remove all the mass-dependent factors when analyzing the clusters from each simulation.
These simulated clusters are not precisely self-similar, in that we typically find a 
$\sim$20\% scatter in entropy at a given scale radius outside $0.1 r_{200}$ and
a somewhat larger variance inside this radius.  However, we find no systematic
trends in the scaled entropy with cluster mass; in other words, we find $K(r/r_{200}) \propto
\Tx$ to within $\sim$20\% across the entire mass range of adequately resolved
clusters.  Thus, our models do not support the suggestion of \citet{vss03} 
that shock heating associated with large-scale structure formation alters 
the $\Lx$-$\Tx$ relation by preferentially elevating the entropy levels in groups.  All of the gravitationally-driven entropy production that happens on large scales is well resolved in the simulations, and the simulated clusters turn out to be self-similar over almost two orders of magnitude in mass.

\subsection{Entropy Scaling Laws}

The temperature of a self-similar cluster depends primarily on the mass $M_{200}$
within the radius $r_{200}$, motivating us to define the characteristic temperature scale
\begin{equation}
 T_{200} \equiv \frac {GM_{200} \mu m_{\rm p}} {2 r_{200}} \; \; .
\end{equation}
Here and throughout the paper we write the temperature in energy units, implicitly 
absorbing Boltzmann's constant into $T$ because observed X-ray temperatures 
are so often quoted in units of keV.  Simulations of non-radiative clusters generally 
find that $\Tx \approx T_{200}$ with a scatter of $\sim$10\% \citep{Voit_RMP} 
that apparently depends on the effective resolution of a simulation and the numerical 
techniques it employs.  

The natural entropy scale in the ICM of a non-radiative cluster is therefore $K_{200} 
\equiv T_{200} \bar{n}_e^{-2/3}$, where $\bar{n}_e$ equals $200 \Omat^{-1}$ 
times the mean electron density of the universe, which would be the mean 
electron density inside $r_{200}$ if the electron to dark-matter
ratio remained constant.  In a $\Lambda$CDM cosmology with a baryon density 
$\Omega_{\rm b} = 0.022 h^{-2}$ 
one finds $\bar{n}_e = (1.45 \times 10^{-4} \, {\rm cm}^{-3})(\Omat/0.3)(1+z)^3$,
giving
\begin{eqnarray}
  K_{200} & =  & 362 \, {\rm keV \, cm^2} \, 
		\frac {\Tx} {1 \, {\rm keV}} \,
		   \left( \frac {T_{200}} {\Tx}  \right)
                    \nonumber \\
           ~ & ~ & \; \; \; \; \; \; \times
                           \left[ \frac {H(z)} {H_0} \right]^{-4/3} \left(  \frac {\Omat} {0.3} \right)^{-4/3} \; \; ,
  \label{eq-ke200}
\end{eqnarray}
where $\Omat$ is the current value of the matter-density parameter.
Writing $K_{200}$ in this way makes explicit the fact that the observed temperature of a 
cluster is not necessarily a reliable guide to the underlying value of $K_{200}$.  If the ICM of a 
real cluster is either hotter or cooler than $T_{200}$, the characteristic temperature
of its halo, then one must apply the correction factor $T_{200}/\Tx$ when computing
the value of $K_{200}$.

Radiative cooling introduces another entropy scale into the ICM that does not enter 
into the comparison of simulated clusters presented here but may be tied to the observed 
$K \propto \Tx^{2/3}$ scaling in real clusters.  Gas of temperature $T$ emitting pure
thermal bremsstrahlung radiation radiates an energy equivalent to its thermal energy
in a time period $t$ if its specific entropy is
\begin{equation}
 K_c \approx 81 \, {\rm keV \, cm^2} \,  
                     \left( \frac {T} {1  \, {\rm keV}}   \right)^{2/3} 
                     \left( \frac{t} {14 \, {\rm Gyr}} \right)^{2/3}\; \; .
\end{equation}
Because this entropy threshold for cooling is quite similar to the observed core entropies
of many clusters, it seems a quite natural explanation for the $K_{0.1} \propto \Tx^{2/3}$
scaling found in cluster cores \citep{vp03}.  However, it is less clear why 
$K \propto \Tx^{2/3}$ should hold when $K \gg K_c$.

\begin{figure}
\includegraphics[width=84mm]{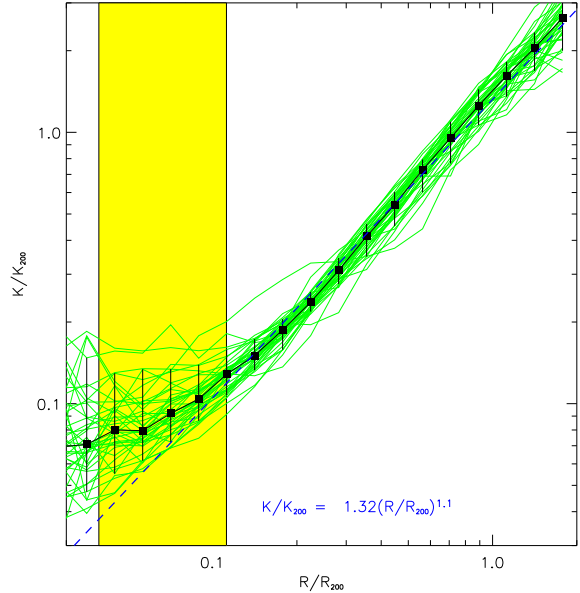}
\caption{ 
Dimensionless entropy $K/K_{200}$ as a function of scale radius $r/r_{200}$ for 40 clusters
simulated with the SPH code GADGET.  Black squares show the median profile, and the 
dashed line illustrates the power-law relation $K/K_{200} = 1.32 (r/r_{200})^{1.1}$.  Most 
of the entropy profiles shown lie close to this relation in the radial range 
$0.1 \lesssim r/r_{200} \lesssim 1.0$.  At smaller radii, the entropy profiles generally
flatten, and their dispersion increases.  The shaded box shows the range of radii over
which the gravity begins to depart from a precise inverse square law because of gravitational softening.
Even though the point at which the entropy profiles begin to flatten coincides with the 
outer edge of this box, we suspect that the flattening is real because the better-resolved, 
higher-mass clusters show the same amount of flattening as the lower-mass clusters when 
scaled relative to $r_{200}$. 
\label{kr_sph}}
\end{figure}

\begin{figure}
\includegraphics[width=84mm]{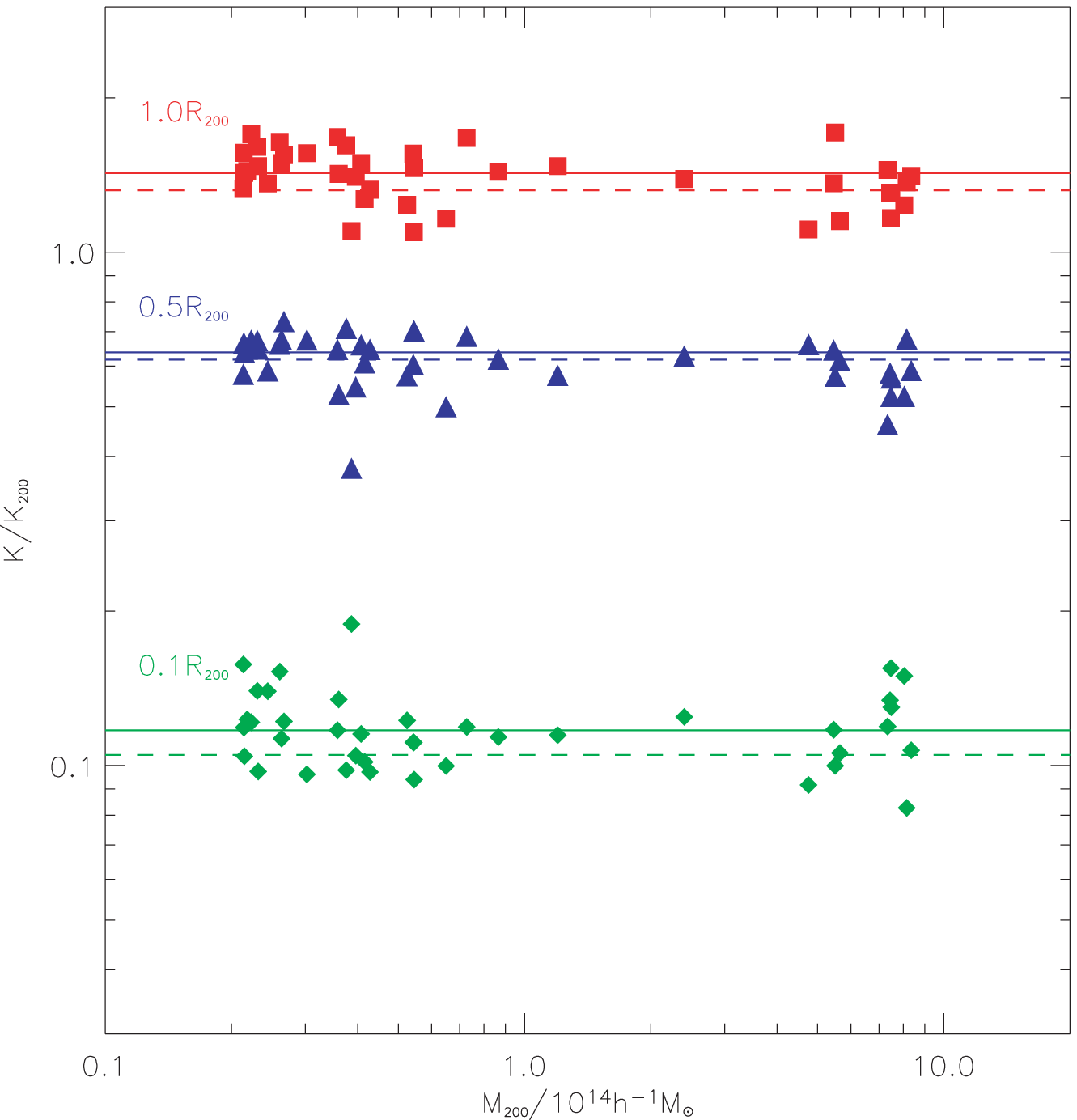}
\caption{
Dimensionless entropy $K/K_{200}$ as a function of halo mass $M_{200}$ at the scale
radii $r/r_{200} = 0.1$, 0.5, and 1.0 in non-radiative clusters simulated with the SPH code 
GADGET.   Squares show entropy at $r_{200}$, triangles show entropy at $0.5 r_{200}$,
and diamonds show entropy at $0.1 r_{200}$.  Solid lines give the median values of
$K/K_{200}$ at each radius, and dashed lines give the corresponding values from the
power-law relation shown in Figure~\ref{kr_sph}.  No systematic trends with mass are evident.
\label{mvar_sph}}
\end{figure}

\subsection{SPH Simulations}

The first set of simulated non-radiative clusters we will consider was produced by the
entropy-conserving version of the smoothed-particle hydrodynamics code GADGET 
\citep{syw01, sh02_entcons}  with $\Omat = 0.3$, $\Olam = 0.7$, 
$\Obar = 0.045$, $h = 0.7$, and $\sigma_8 = 0.9$.  Most of this set comes from 
the non-radiative simulation described in \citet{Kay04}, from which we take the thirty
most massive clusters, ranging from  $2.1 \times 10^{13} \, h^{-1} \, M_\odot$ to 
$7.5 \times 10^{14} \, h^{-1} \, M_\odot$.  These clusters were modeled within a box 
of comoving length $60 \, h^{-1} \, {\rm Mpc}$ using an equivalent Plummer
softening length of $20 \, h^{-1} \, {\rm kpc}$ in comoving coordinates.  To supplement
the high-mass end of the set, we also include the ten high-mass clusters simulated
by \citet{ktjp04}, using the same code and the same cosmological parameters.  
These clusters, which were extracted from a much larger simulation volume and 
then individually resimulated at a spatial resolution comparable to that of the 
$60 \, h^{-1} \, M_\odot$ simulation, range from 
$5.5 \times 10^{14} \, h^{-1} \, M_\odot$ to $8.4 \times 10^{14} \, h^{-1} \, M_\odot$.  
Thus, all forty clusters in this overall sample are reasonably well-resolved.

Figure~\ref{kr_sph} shows the dimensionless entropy profiles of these clusters, averaged over
radial bins.  The average entropy in each spherical shell is defined to be the mean temperature 
in that shell divided by the two-thirds power of the mean electron density within the shell. 
Most of the spherically-averaged profiles are virtually identical at $r > 0.1 r_{200}$, 
consistent with the expectation of self-similarity.  Within the 
range $0.2 \lesssim r/r_{200} \lesssim 1.0$, the power law $K(r) \propto r^{1.1}$ shown by
the dashed line is a good approximation, in agreement with the spherical-accretion
models of \citet{tn01} and \citet{vbblb03} and the simulations of
\citet{borg02}.  Inside of $0.1 r_{200}$ there is more diversity.  Some of the
simulated clusters have nearly isentropic cores, while others do not.  The flattening of
the entropy profiles within the core is likely to be a real effect because the degree of 
flattening does not depend on cluster mass.  If the flattening were due to a resolution effect, 
then it would be more pronounced in smaller, lower-mass clusters, whose physical size 
is smaller relative to the resolution length of the simulation.  
However, the same kind of flattening is seen in
the better-resolved, higher-mass clusters simulated by \citet{ktjp04}.

Figure~\ref{mvar_sph} shows that dimensionless entropy measured at a fixed
scale radius does not depend on halo mass.  No significant deviations from the
approximate self-similar profile $K(r) = 1.32\,K_{200}\,(r/r_{200})^{1.1}$ are seen over 
the entire mass range, from $2 \times 10^{13} \, h^{-1} \, M_\odot$ to $8 \times
10^{14} \, h^{-1} \, M_\odot$.  We have assessed the scatter in dimensionless
entropy with the quantity 
\begin{equation}
 \frac {\Delta K} {K} \equiv \frac {K_{90\%} - K_{10\%}} {2 K_{\rm 50\%}} \; \; ,
\end{equation}
where $K_{X\%}$ is the $X$th percentile of the dimensionless entropy
at a given radius.  The average value of $\Delta K/K$ in the range
$0.2 < r/r_{200} < 1.0$ is 0.12, and over the larger range $0.02 < r/r_{200} <
1.8$ its value is 0.28.

\subsection{AMR Simulations}

The other set of clusters we will consider was produced by the adaptive-mesh refinement
code ENZO \citep{Bryan99, nb99, Oshea_ENZO} with $\Omat = 0.3$, $\Olam = 0.7$, 
$\Obar = 0.04$, $h = 0.67$, and $\sigma_8 = 0.9$.  Twenty-one of these non-radiative clusters 
were modeled within a box of comoving length $50 \, h^{-1} \, {\rm Mpc}$ using  
mesh refinement to produce an effective resolution of $20 \, h^{-1} \, {\rm kpc}$ in 
comoving coordinates.  Further details about this simulation are given in \citet{bv01}.  
To populate the upper end of the mass range, we added ten more clusters
to our sample, drawn from the simulation described in \citet{Loken02}, with a box size
of $256 \, h^{-1} \, {\rm Mpc}$ and an effective resolution of $15 \, h^{-1} \, {\rm kpc}$.
That larger simulation assumed a slightly different cosmology, with $\Omat = 0.3$, $\Olam = 0.7$, 
$\Obar = 0.026$, $h = 0.7$, and $\sigma_8 = 0.928$.  However, dividing by the
appropriate value of $K_{200}$ when constructing the dimensionless entropy
profiles compensates for the differing baryon density scale. 
The overall AMR cluster set ranges in mass from $2.7 \times 10^{13} \, h^{-1} \, M_\odot$ to 
$1.4  \times 10^{15} \, h^{-1} \, M_\odot$.

\begin{figure}
\includegraphics[width=84mm, trim = 1.0in 1.2in 0.9in 1.0in , clip]
{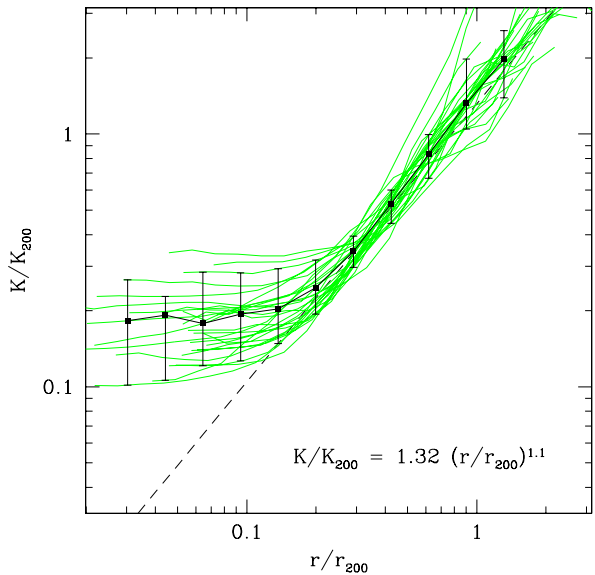}
\caption{ 
Dimensionless entropy $K/K_{200}$ as a function of scale radius $r/r_{200}$ for 31 clusters
simulated with the AMR code ENZO.  As in Figure~\ref{kr_sph}, most 
of the entropy profiles shown lie close to the relation $K/K_{200} = 1.32 (r/r_{200})^{1.1}$
in the radial range $0.2 \lesssim r/r_{200} \lesssim 1.0$.  However, the flattening at smaller radii is 
more pronounced than in the SPH simulation, leading to substantially higher entropy
levels near the origin.  
\label{kr_amr}}
\end{figure}

\begin{figure}
\includegraphics[width=84mm, trim = 1.0in 1.2in 0.9in 1.0in , clip]
{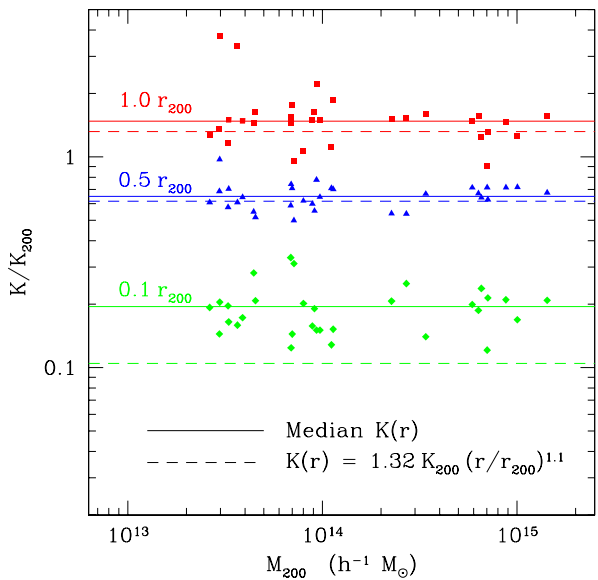}
\caption{
Dimensionless entropy $K/K_{200}$ as a function of halo mass $M_{200}$ at the scale
radii $r/r_{200} = 0.1$, 0.5, and 1.0 in non-radiative clusters simulated with the AMR code 
ENZO.  Squares show entropy at $r_{200}$, triangles show entropy at $0.5 r_{200}$,
and diamonds show entropy at $0.1 r_{200}$.  Solid lines give the median values of
$K/K_{200}$ at each radius, and dashed lines give the corresponding values from the
power-law relation shown in Figure~\ref{kr_sph}.  As in Figure~\ref{mvar_sph}, there are
no discernible systematic trends with mass. 
\label{mvar_amr}}
\end{figure}

As can be seen in Figure~\ref{kr_amr}, the entropy profiles measured in this sample of simulated 
clusters are also nearly self-similar.  Beyond the scale radius $0.2 r_{200}$, the power-law slope 
and normalization of the dimensionless entropy profiles is quite close to those found in the SPH 
clusters, as illlustrated by the dashed line.   However, the AMR profiles within this radius are 
significantly flatter than the SPH profiles (\S~\ref{sec-coreK}).

Likewise, the dimensionless entropy profiles of our AMR clusters also show no systematic trends 
with mass.  Figure~\ref{mvar_amr} samples the dimensionless entropy in this set of clusters at the
same three scale radii as in Figure~\ref{mvar_sph}.  Again, the points cluster around the median
profile, with no dependence on $M_{200}$.  The scatter in entropy likewise has properties 
similar to those found in the SPH simulation:  $\Delta K / K = 0.21$ over the range $ 0.2 \lesssim 
r/r_{200} \lesssim 1.0$ and $\Delta K / K = 0.31$ over the complete range.

\section{Cross-Comparison of Simulations}

It should already be clear from the preceding section that the median
dimensionless entropy profiles found in both cluster simulations are
nearly the same outside of $0.2 r_{200}$.  Here we discuss some of
the systematic differences between the two sets of simulated clusters
and then provide a simple analytical formula for the median entropy
profile that is consistent with both samples.  To help observers use this
formula, we conclude this section with a brief comparison to actual data,
including a discussion of how offsets of $\Tx$ with respect to 
$T_{200}$ affect the determination of the baseline 
profile for an observed cluster.  

\subsection{Entropy within the Core}
\label{sec-coreK}

\begin{figure}
\includegraphics[width=84mm, trim = 1.0in 1.2in 0.9in 1.0in , clip] 
{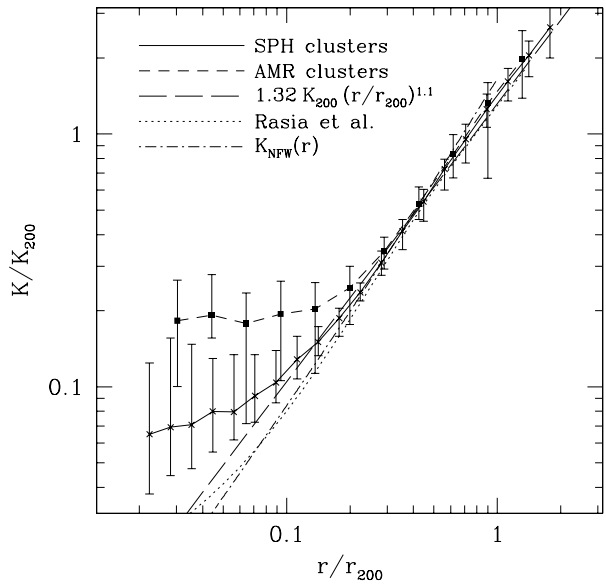}
\caption{   
Median entropy profiles from cluster simulations without
cooling or non-gravitational heating. 
A solid line connecting crosses shows the median profile for the SPH clusters.
A dashed line connecting squares shows the median profile for the AMR clusters.
Error bars give the 10\% percentile to 90\% percentile range.  The long dashed 
line illustrates the power-law approximation $K(r) = 1.32\,K_{200}\,(r/r_{200})^{1.1}$.
The dotted line gives an analytical entropy profile derived from simulations
by \citet{rtm03}.  The dot-dashed line shows the profile $K_{\rm NFW}(r)$ corresponding
to an NFW gas-density profile of concentration $c=5$ that is in hydrostatic equilibrium 
within a dark-matter density profile of identical shape \citep{vbbb02}.  Notice that
both of the median profiles agree very well in the range $0.2 < r/r_{200} < 1.0$ and
that the analytical approximations accurately represent the median profiles in this
range.  However, the AMR and SPH median profiles differ by as much as a factor 
of two within the cluster core.
\label{kprof_comp}}
\end{figure}

The median dimensionless entropy profiles of clusters in our two simulation 
sets agree well outside the cluster core but disagree within the cluster core.  
Figure~\ref{kprof_comp}  illustrates the discrepancy.  This type of discrepancy
between SPH and AMR is nothing new.  It was previously hinted at in the Santa
Barbara cluster comparison \citep{f99_sb}, but that result was not definitive because 
the sample size was a single cluster.  Here we confirm it for a large sample of clusters
with a range of masses and simulated with substantially higher spatial and mass
resolution.  Thus, we do not attempt to fit an analytical form to our median profiles 
within $0.1 r_{200}$;  the numerical techniques used to model this region do not yet 
give a reliable answer.

Because the primary purpose of this paper is to provide a baseline profile for observers
to use outside the cluster core, we leave a detailed analysis of the reasons for
this entropy discrepancy for future work.  It is an important problem to pursue because
of its implications for cooling and condensation of gas within cluster cores.
Larger amounts of entropy production within the core, as in the AMR code, 
will more effectively inhibit cooling there, perhaps mitigating the ``cooling-flow
problem" in clusters of galaxies \citep[see][for a recent review]{dv04}.

 \subsection{Power-Law Approximations}

Previous theoretical work has shown that the entropy profiles of non-radiative clusters
approximately follow a power law with $K(r) \propto r^{1.1}$ \citep[e.g.,][]{tn01, borg02, vbblb03}, 
but these efforts have not provided a normalization for that power-law profile 
in a form that is useful to observers.  Here we rectify that situation.   If we fix the power-law
slope of the entropy profile at 1.1 and fit the SPH clusters in the radial range 
$0.2 \leq r/r_{200} \leq 1.0$, we find
\begin{equation}
 K(r) = 1.32 \pm 0.03 \,  K_{200} \, (r/r_{200})^{1.1} \; \; .
 \label{eq_sph11}
\end{equation}
Doing the same for the AMR clusters yields
\begin{equation}
 K(r) = 1.41 \pm 0.03 \,  K_{200} \, (r/r_{200})^{1.1} \; \; ,
\end{equation}
a normalization just slightly higher than that for the SPH clusters.

Figure \ref{kprof_comp} compares the power-law fit from equation~(\ref{eq_sph11})
with the median profiles from both simulations and with two other analytical entropy 
profiles.  One of the analytical profiles is constructed from the analytical temperature
and density profiles developed by \citet{rtm03} to fit their SPH models of non-radiative
clusters.  The other is the entropy profile of intracluster gas with an NFW density distribution
with concentration $c=5$ assuming that it is in hydrostatic 
equilibrium within a dark-matter halo whose density
distribution has the same shape \citep{vb01, vbbb02}.  All of these profiles agree very well
with the power-law profile of equation~(\ref{eq_sph11}) in the range $0.2 \leq r/r_{200} \leq
1.0$.  Also, the power-law profile remains a good representation of our SPH simulations
down to $\approx 0.1\,r_{200}$.  We note, however, that the \citet{rtm03} profile shows 
much less flattening in the core than the simulations analyzed in this paper, 
perhaps because it is based on an earlier version of
GADGET that does not explicitly conserve entropy \citep[see also][]{Ascasibar03}.
  
\begin{figure}
\includegraphics[width=84mm, trim = 1.0in 1.2in 0.9in 1.0in , clip] 
{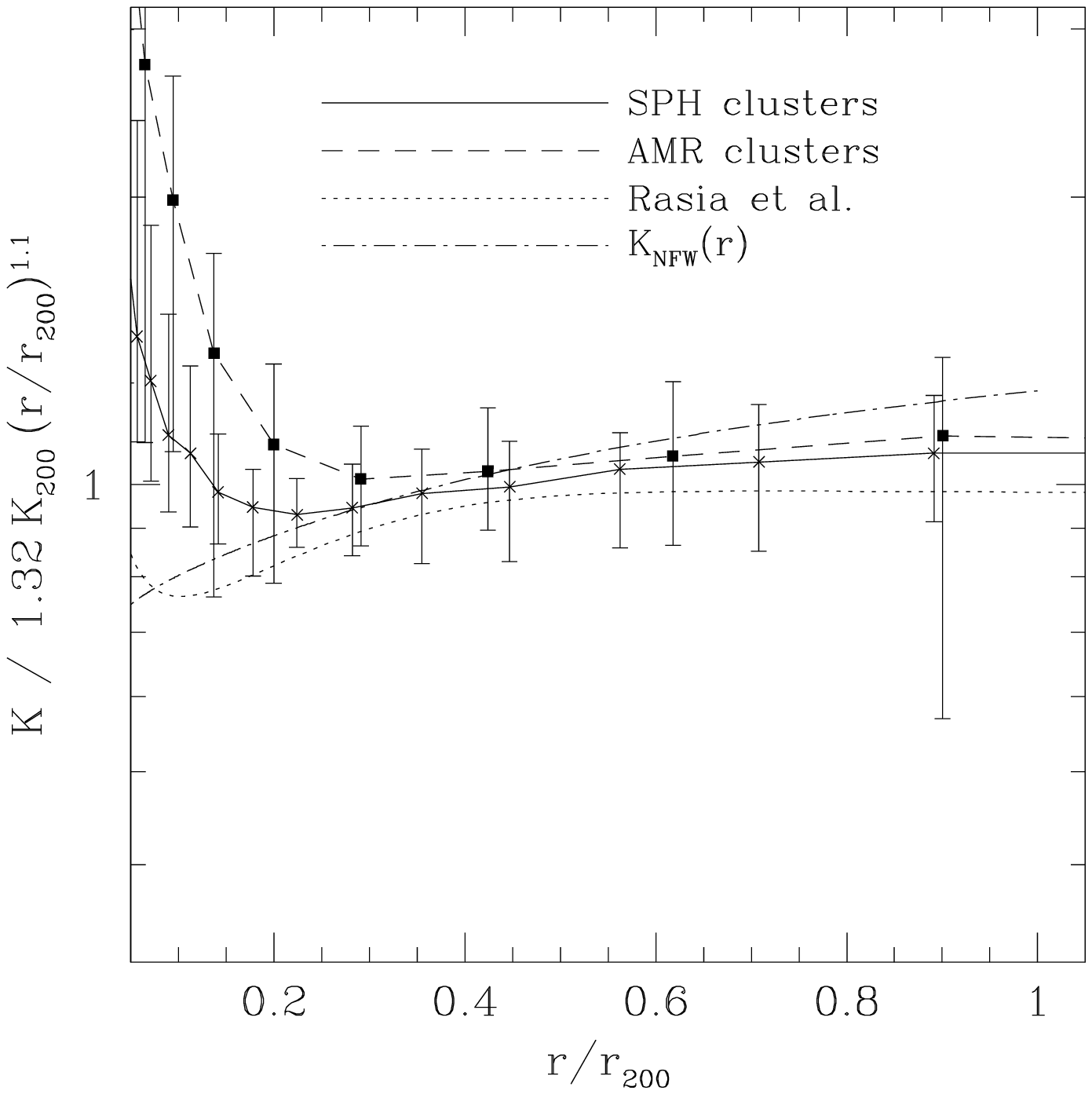}
\caption{ 
Median dimensionless entropy profiles from simulations, divided by the
power-law profile $K(r) = 1.32\,K_{200}\,(r/r_{200})^{1.1}$.  Crosses connected
by a solid line show the median profile from the SPH simulation, and squares connected
by a dashed line show the median profile from the AMR simulation.  
Dashed and dot-dashed lines show the \citet{rtm03} and NFW-like profiles, respectively.
A slight rise in the median points as radius increases beyond $0.2r_{200}$ suggests
that the assumed power-law index of 1.1 is slightly too small.
\label{kpl11}}
\end{figure}

\begin{figure}
\includegraphics[width=84mm, trim = 1.0in 1.2in 0.9in 1.0in , clip] 
{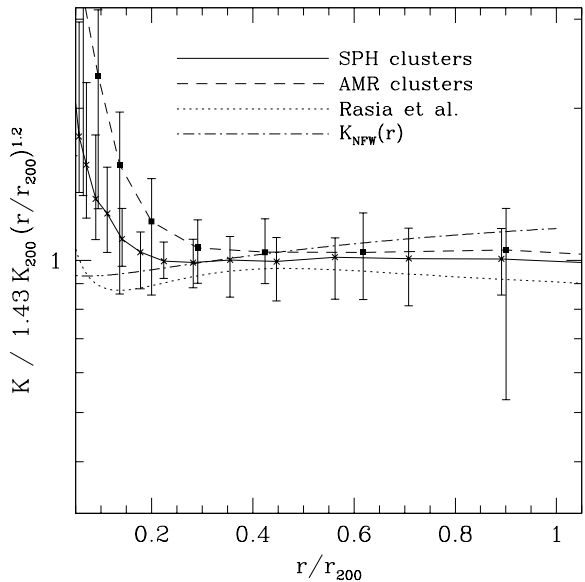}
\caption{
Median dimensionless entropy profiles from simulations, divided by the
power-law profile $K(r) = 1.43\,K_{200}\,(r/r_{200})^{1.2}$.  Crosses connected
by a solid line show the median profile from the SPH simulation, and squares connected
by a dashed line show the median profile from the AMR simulation.  
Dashed and dot-dashed lines show the \citet{rtm03} and NFW-like profiles, respectively.
A power-law index of 1.2 seems to be a better description of the entropy profile
beyond $0.2r_{200}$ than the standard index of 1.1.
\label{kpl12}}
\end{figure}

Up to this point, we have been fitting the median profiles with a $K(r) \propto r^{1.1}$
power law because that is the standard power-law index in the literature, and it
appears to be consistent with the highest quality cluster observations 
\citep[e.g.,][]{pa03}.  However, Figure~\ref{kpl11} suggests that a power-law index
of 1.1 might be slightly too shallow to be the best representation of the median
profiles outside of $0.2 r_{200}$.  Fitting a power-law of index 1.2 to the median
profiles gives the following results:
\begin{equation}
 K(r) = 1.43 \pm 0.01 \,  K_{200} \, (r/r_{200})^{1.2} \; \; .
 \label{eq_sph12}
\end{equation}
for the SPH clusters and
\begin{equation}
 K(r) = 1.48 \pm 0.02 \,  K_{200} \, (r/r_{200})^{1.2} \; \; 
\end{equation}
for the AMR clusters.  Figure~\ref{kpl12} shows that dividing the median profiles
by equation~(\ref{eq_sph12}) makes the resulting profiles flatter, indicating
that $K(r) \propto r^{1.2}$ is a better description of the outer parts of non-radiative
clusters.  Indeed, if we allow the power-law index to be a free parameter and
fit the median profiles in the range $0.2 \leq r/r_{200} \leq 1.0$, we find
\begin{equation}
 K(r) = 1.45 \pm 0.01  \,  K_{200} \, (r/r_{200})^{1.21 \pm 0.01} \; \; .
\end{equation}
for the SPH clusters and
\begin{equation}
 K(r) = 1.51 \pm 0.03 \,  K_{200} \, (r/r_{200})^{1.24 \pm 0.03} \; \; 
\end{equation}
for the AMR clusters.  In all of these fits the error bars correspond to 1$\sigma$.

 \subsection{Applications of the Baseline Profiles}

\begin{figure}
\includegraphics[width=84mm, trim = 0.8in 1.2in 0.9in 1.0in , clip] 
{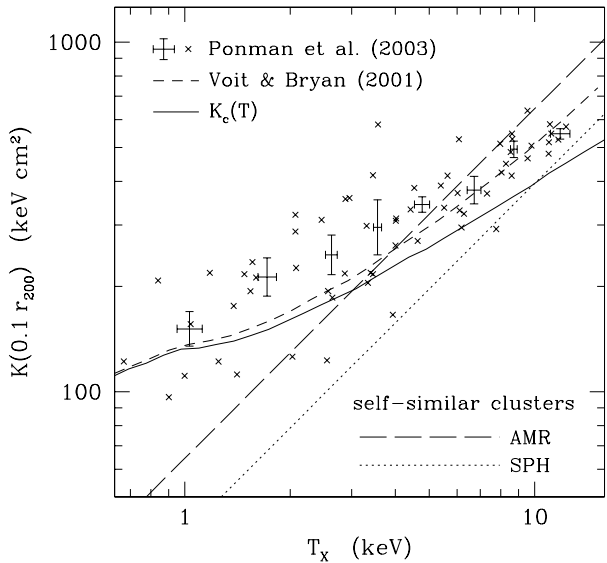}
\caption{ 
Relationship between core entropy and the cooling threshold.  
Each point with error bars shows the mean core entropy $K_{0.1}$, 
measured at $0.1\,r_{200}$, for eight clusters within a given 
temperature bin, and small 
crosses show measurements for individual clusters \citep{psf03}.
The long-dashed and dotted lines show self-similar relations calibrated using 
the median values of $K_{0.1}$ derived from our AMR and SPH simulations,
respectively, assuming that $\Tx = T_{200}$.  High-temperature clusters
appear to be consistent with the median self-similar profiles, but the trend
to lower-temperatures more closely tracks the cooling threshold $K_c(T)$
(solid line), defined to be the entropy at which the cooling time equals 14~Gyr
\citep{vp03}.  The short-dashed line shows the predicted entropy at $0.1r_{200}$ 
in the model of \citet{vb01}.  
\label{ent_core}}
\end{figure}

The entropy profiles computed here for non-radiative clusters provide a baseline
for assessing the impact of non-gravitational processes on the intracluster medium.
However, in order to compare the entropy profiles of a real cluster to these self-similar
baselines, one needs to know the value of $K_{200}$ for the cluster.  This characteristic
entropy scale can be simply computed from equations (1) and (2) if the cluster mass
$M_{200}$ has been accurately measured.  Otherwise, one must infer $K_{200}$ from
$\Tx$ and a relation between $T_{200}$ and $\Tx$ or, equivalently, a relation
between $M_{200}$ and $\Tx$.
Figure~\ref{ent_core} compares entropy values measured at $0.1 r_{200}$ with those
predicted for self-similar clusters.  The long-dashed and dotted lines show the predictions
for non-radiative AMR and SPH clusters, respectively.  Here we set $\Tx = T_{200}$ because
any offset between $\Tx$ and $T_{200}$ is small compared with the difference between
the two simulation sets.  As shown by \citet{psf03} and \citet{vp03}, the measurements clearly
do not agree with the self-similar models, which predict that $K(0.1 r_{200}) \propto \Tx$.
Instead, the measurements track the cooling threshold $K_c(T) \propto \Tx^{2/3}$.
We wish to point out, however, that the hot clusters are consistent with non-radiative models
at this radius.  This finding contrasts with Figure~1 of \citet{vp03}, in which the locus for 
self-similar clusters is mistakenly a factor of two too low, owing to a unit conversion error.

\begin{figure}
\includegraphics[width=84mm, trim = 0.8in 1.2in 0.9in 1.0in , clip] 
{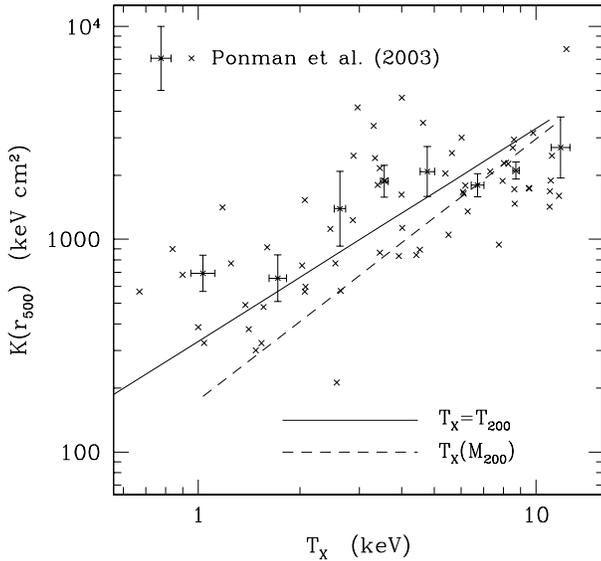}
\caption{
Entropy at $r_{500}$ as a function of 
cluster temperature.  Each point with error bars shows the mean
value of $K(r_{500})$ implied by the density and temperature profiles
of eight clusters within that temperature bin \citep{psf03}, and small crosses
show measurements for individual clusters.
The solid line shows the median entropy at $r_{500}$ for self-similar
clusters, assuming that $\Tx = T_{200}$.  The dashed line shows how the
mapping of this median entropy onto $\Tx$ changes when the $\Tx(M_{200})$
relationship observed by \citet{spflm03} is used to determine $\Tx$. 
The most massive clusters are consistent with the self-similar clusters modeled 
without non-gravitational processes when this observational $M_{200}$-$\Tx$ 
relation is used. 
\label{ent_excess}}
\end{figure}

Figure~\ref{ent_excess} compares entropy values measured at $r_{500} \approx 0.66 r_{200}$, 
within which the mean mass density is 500 times the critical density, with the baseline profiles.
The solid line shows the baseline entropy level derived assuming $\Tx = T_{200}$, which
slightly exceeds the measured entropy levels at this radius in hot clusters.  This apparent shortfall
in the observed entropy levels of hot clusters goes away when we account for the difference between 
$\Tx$ and $T_{200}$.  A dashed line shows the baseline entropy level at $r_{500}$ computed 
using values of $K_{200}$ derived from the $M_{200}$-$\Tx$ relation of \citet{spflm03}, and this
line is consistent with the entropy measurements at $r_{500}$ in hot clusters.  This consistency again 
contrasts with the results of \citet{vp03}, in which the locus for self-similar 
clusters was mistakenly placed too low.  However, clusters below about 6~keV still show 
a clear entropy excess, which is even more pronounced when the observed 
$M_{200}$-$\Tx$ relation is used to compute $K_{200}$.

\section{Summary}

Our intention in this paper has been to provide a simple analytical form for the entropy profiles
of non-radiative clusters to use as a baseline when trying to measure the impact of
non-gravitational processes on the intracluster medium.  To that end, we analyzed two
different sets of simulated clusters, one created with a Lagrangian SPH code
and the other with an Eulerian AMR code.  Thirty to forty entropy profiles were produced
by each code and these profiles were found to be approximately self-similar, 
with the $K(r/r_{200}) \propto \Tx$ scaling expected of non-radiative clusters.  
The simulated profiles depend very little on halo mass once the expected scaling
is divided out.  This result confirms that non-gravitational processes are necessary 
to produce the observed scaling relations of clusters. 

The median entropy profiles from the two simulations agree to within 7\% outside 
of $0.2 r_{200}$ but disagree in the cluster core.  In the outer parts of clusters
the power-law profile $K(r) = 1.32\,K_{200}\,(r/r_{200})^{1.1}$ is a good representation
of the baseline profile expected in the non-radiative case.  However, our results suggest
that the baseline profile in the radial range $0.2 \leq r/r_{200} \leq 1.0$ is better fit by
a $K(r) \propto r^{1.2}$ power law, rather than the standard $K(r) \propto r^{1.1}$ law
found by \citet{tn01}.  Inside of $0.2 r_{200}$ the discrepancy between the AMR clusters
and the SPH clusters is quite substantial.  Tracking down the origin of this discrepancy
is important, because radiative cooling rates in clusters depend on the core entropy 
level.  

Our comparison between the self-similar entropy profiles derived from these simulated 
clusters with measurements of entropy at $0.1\, r_{200}$ and $r_{500}$ in real clusters 
updates and corrects the findings of \citet{vp03}.  There is a clear entropy excess
in cool clusters, presumably stemming from non-gravitational processes.  However, 
clusters hotter than $\sim 6 \, \keV$ appear to converge to the self-similar profile 
at radii $\gtrsim 0.1 \, r_{200}$.

\vspace*{1em}
We thank Trevor Ponman for persistently asking the questions that motivated this 
project and and Volker Springel for generously allowing us to use 
GADGET2 before its public release. 
The SPH simulation data were generated using COSMA, the 670-processor 
COSmology MAchine at the Institute for Computational Cosmology in Durham, 
as part of the programme of the Virgo Supercomputing Consortium.   
GMV received support for this project from NASA's Astrophysics Theory program 
through grant NNG04GI89G.
STK is supported by the UK Particle Physics and Astronomy 
Research Council (PPARC).  GLB thanks PPARC and the Leverhulme 
Trust for support.


\end{document}